\documentclass[12pt]{iopart}

\usepackage{graphicx}
\usepackage{setstack} 
\usepackage{iopams}
\usepackage{color}
\usepackage{amssymb}
\usepackage{mathrsfs}

\begin{document}

\title[Speeding up the first-passage for subdiffusion]{Speeding up the
first-passage for subdiffusion by introducing a finite potential barrier}

\author{Vladimir V. Palyulin$^{\ddagger,\dagger}$, and Ralf
Metzler$^{\ddagger,\sharp}$}
\address{$\ddagger$ Institute for Physics \& Astronomy, University of Potsdam,
D-14476 Potsdam, Germany\\
$\dagger$ Physics Department, Technical University of Munich,
D-85747 Garching, Germany\\
$\sharp$ Physics Department, Tampere University of Technology, FI-33101
Tampere, Finland}

\date{\today}

\begin{abstract}
We show that for a subdiffusive continuous time random walk with scale-free
waiting time distribution the first-passage dynamics on a finite interval
can be optimised by introduction of a piecewise linear potential barrier.
Analytical results for the survival probability and first-passage density
based on the fractional Fokker-Planck equation are shown to agree well with
Monte Carlo simulations results. As an application we discus an improved
design for efficient translocation of gradient copolymers compared to
homopolymer translocation in a quasi-equilibrium approximation.
\end{abstract}


\section{Introduction}

The first passage of a stochastic process across a certain, pre-set value renders
vital information on the underlying dynamics \cite{Redner}. Thus, it quantifies
how long it takes a share to cross a given price threshold in the stock exchange.
One of the famed historical versions of such a first passage problem is the
Pascal-Huygens gambler's ruin, i.e., the number of rounds of a game it takes until
the first gambler goes broke. For a particle diffusing in space, one is normally
interested in the time it takes the particle to reach a given position after its
initial release at some other position. Here we pursue the question of how we can
optimise the first passage of a particle from point O to X, when the values
of the potential $U(x)$ is different at these two points. It was previously shown
that for a Brownian particle the mean first passage time can be significantly
reduced in a piecewise linear potential when the particle first has to cross a
large potential barrier located close to its starting point and in exchange
experiences a drift towards the target for the remaining part of its trajectory
\cite{mfpt_min}.

Can similar effects be observed when instead of a Brownian particle we consider a
particle in a strongly disordered environment? To answer this question we study a
particle that performs anomalous diffusion \cite{report}
\begin{equation}
\label{msd}
\langle x^2(t)\rangle\simeq K_{\alpha}t^{\alpha}
\end{equation}
with anomalous diffusion exponent $0<\alpha<1$ and the generalised diffusion
coefficient $K_{\alpha}$ of physical dimension $[K_{\alpha}]=\mathrm{cm}^2/\mathrm{
sec}^{\alpha}$. Microscopically, we assume that the particle follows a
Scher-Montroll continuous time random walk (CTRW), in which successive jumps of the
particle are separated by independent, random waiting times $\tau$ with power-law
distribution,
\begin{equation}
\label{wtd}
\psi(\tau)\sim\frac{1}{\tau^{\star}}\left(\frac{\tau^{\star}}{\tau}\right)^{1+
\alpha},
\end{equation}
where $\tau^{\star}$ is a scaling factor of physical dimension of time, such that
no characteristic time scale $\langle\tau\rangle$ exists \cite{montroll,scher,igor}.
Realisations of CTRW subdiffusion were observed in a variety of systems including
charge  carrier diffusion in amorphous semiconductors \cite{scher},
the motion of submicron particles in living biological cells \cite{tabei}, the
dynamics of tracer beads in an actin mesh \cite{wong}, or the motion of
functionalised colloidal particles along a complementary, functionalised surface
\cite{xu}.

As we show here based on analytical calculations and numerical analyses the
introduction of a piecewise linear potential indeed renders the first passage
behaviour of subdiffusive processes more efficient. This is demonstrated in
terms of the density of first passage times and the associated survival
probability, as well as a recently defined efficiency parameter. We discuss
potential applications of our findings to the translocation of polymers through
narrow channels.

\section{Model and Analytical Results}

\begin{figure}
\begin{center}
\includegraphics[width=8.6 cm]{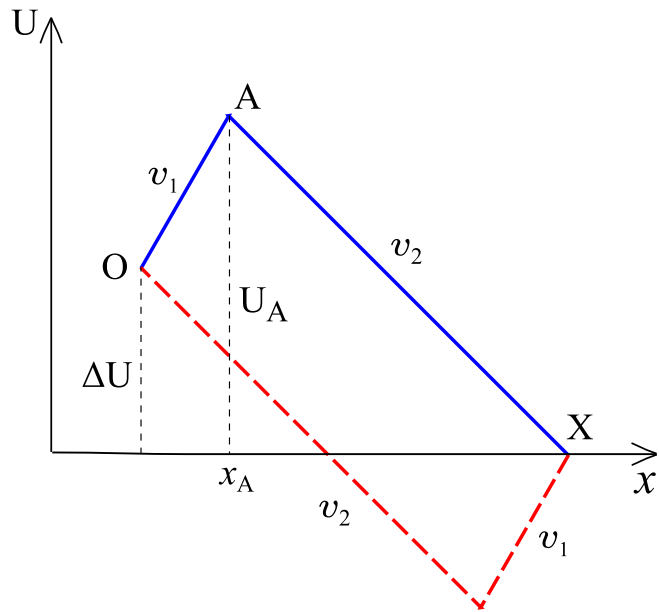}
\end{center}
\caption{Sketch of a piecewise linear potential between the initial particle
position in point O and the target point X. Initially the particle needs to
cross the potential barrier culminating in A fuelled by thermal fluctuations,
before a constant downward slope pushes the particle towards X. The same result
is obtained by first sliding down and then crossing the barrier (red dashed
lines).}
\label{Sketch}
\end{figure}

We assume that the particle starts at point O and diffuses until it reaches the
point X located at $x_X=1$ in normalised units. On its way it passes through a
piecewise linear potential with a change of slope at point A at $x_{\mathrm{A}}$
(see Fig.~\ref{Sketch}). We denote the values of potential at these points as $U_{
\mathrm{O}}$, $U_{\mathrm{A}}$, and $U_{\mathrm{X}}=0$. Thus, by help of thermal
fluctuations the particle first
crosses the potential maximum at point A before being advected towards the
target X for the rest of the way. At the starting point O a reflecting boundary
condition is assumed, while at the target we implement an absorbing boundary to
account for the first passage problem \cite{Redner}.

The basis for the analytical description of this subdiffusion problem with given
distribution (\ref{wtd}) of waiting times $\tau$ in the long-time limit $t\gg\tau
^{\star}$ is given by the fractional Fokker-Planck equation \cite{report} which
we here write in the integral form
\begin{equation}
P(x,t)-P(x,0)=\,_0D^{-\alpha}_t\left(\frac{\partial}{\partial x}\frac{U'(x)}
{m\eta_{\alpha}}+K_{\alpha}\frac{\partial^2}{\partial x^2}\right)P(x,t)
\label{fpe}
\end{equation}
where $P(x,0)$ is the initial condition, $U'(x)$ is the derivative of the external
potential, $m$ is the particle mass, and $\eta_{\alpha}$ the friction experienced
by the particle. The Riemann-Liouville fractional integral is defined in terms of
\begin{equation}
_0D^{-\alpha}_tP(x,t)=\frac{1}{\Gamma(\alpha)}\int_0^t\frac{P(x,t')}{(t-t')^{1-
\alpha}}dt',
\end{equation}
representing a Laplace convolution. In the Brownian limit $\alpha=1$,
Eq.~(\ref{fpe}) reduces to the regular Fokker-Planck equation.

For segments with a linear potential in our piecewise linear form $U=m\eta_{\alpha}
v_i$ the fractional Fokker-Planck equation reduces to
\begin{eqnarray}
P_i(x,t)-P_i(x,0)=\,_0D^{-\alpha}_t\left(-v_i\frac{\partial P_i(x,t)}{\partial x}
+K_\alpha\frac{\partial^2P_i(x,t)}{\partial x^2}\right),
\label{ffpe}
\end{eqnarray}
where $i=1,2$ corresponds to the two different slopes of the piecewise linear
potential. In our choice of $U$ the $v_i$ correspond to drift velocities, as
the dimension of $\eta_{\alpha}$ is that of $[\eta_{\alpha}]=\mathrm{sec}^{\alpha
-2}$ \cite{report}. The solution of this equation with a reflective boundary
condition at one end and an absorbing boundary at the other can be found by
methods similar to the Brownian case, compare Ref.~\cite{Redner}. If both boundary
and initial conditions are set as vanishing probability at $x_X=1$, $P(1,t)=0$, for
the absorbing boundary, the flux condition at the origin $x_{\mathrm{O}}=0$ of the
form $j(0,t)=\frac{\partial P(0,t)}{\partial x}-vP(0,t)=\delta(t)$, and the initial
condition $P(x,0)=0$ \cite{Redner}, then
\begin{eqnarray}
P_i(x,t)=\,_0D^{-\alpha}_t\left(K_\alpha\frac{\partial^2 P_i(x,t)}{\partial x^2}-
v_i\frac{\partial P_i(x,t)}{\partial x}\right).
\label{FFPE1}
\end{eqnarray}
Applying the Laplace transform
\begin{equation}
P(x,s)=\int_0^{\infty}P(x,t)e^{-st}dt
\end{equation}
to Eq.~(\ref{FFPE1}), we find the ordinary differential equation
\begin{eqnarray}
s^\alpha P_i(x,s)=\left(K_\alpha\frac{\partial^2 P_i(s,t)}{\partial x^2}-v_i
\frac{\partial P(s,t)}{\partial x}\right).
\end{eqnarray}
The solution of this equation has the form $P(x,t)=A_i e^{\alpha_i x}+B_ie^{\beta
_i x}$ with the exponents
\begin{eqnarray}
\nonumber
\alpha_{1,2}&=&\left(v_{1,2}+\sqrt{v_{1,2}^{2}+4K_{\alpha}s^\alpha}\right)/2K_{
\alpha},\\
\beta_{1,2}&=&\left(v_{1,2}-\sqrt{v_{1,2}^{2}+4K_{\alpha}s^\alpha}\right)/2K_{
\alpha}.
\label{exponents}
\end{eqnarray}
The coefficients $A_{1,2}$ and $B_{1,2}$ are determined by the boundary conditions
$P(1,t)=0$ and $j(0,t)=\delta(t)$ as well as by the continuity of flux and
distribution $P$ at $x=x_{\mathrm{A}}$. This produces the following system of linear
equations,
\begin{eqnarray}
\nonumber
&&A_1 v_1+B_1 v_1-K_{\alpha}\alpha_1A_1-K_{\alpha}\beta_1B_1=1\\
\nonumber
&&A_2e^{-\alpha_2}+B_2e^{-\beta_2}=0\\
\nonumber
&&A_1e^{-\alpha_1x_{\mathrm{A}}}+B_1e^{-\beta_1x_{\mathrm{A}}}=A_2e^{-\alpha_2x_{\mathrm{A}}}+B_2e^{-\beta_2x_{\mathrm{A}}}\\
\nonumber
&&A_1v_1e^{-\alpha_1x_{\mathrm{A}}}+B_1v_1e^{-\beta_1x_{\mathrm{A}}}-K_{\alpha}\alpha_1A_1e^{-\alpha_1
x_{\mathrm{A}}}-K_{\alpha}\beta_1B_1e^{-\beta_1x_{\mathrm{A}}}\\
&&\hspace*{0.8cm}
=A_2v_2e^{-\alpha_2x_{\mathrm{A}}}+B_2v_2e^{-\beta_2x_{\mathrm{A}}}-K_{\alpha}\alpha_2A_2e^{-\alpha_2
x_{\mathrm{A}}}-K_{\alpha}\beta_2B_2e^{-\beta_2x_{\mathrm{A}}}.
\label{system}
\end{eqnarray}

Due to the divergence of the characteristic waiting time $\langle\tau\rangle$
for CTRW subdiffusion processes, even in confined geometries no mean first
passage time exists \cite{physica,Lindenberg}. Below we will therefore analyse the average
time for the 50\% or 90\% probability that the particle has arrived in X.
Analytically, the relevant quantity for this type of process is the probability
density of first arrival, $\wp_{\alpha}(t)$, or the cumulative survival probability,
$\mathscr{S}_{\alpha}(t)=\int_{0}^{1}P(x,t)dx$. Both quantities are related through
$\wp_{\alpha}(t)=-d\mathscr{S}_{\alpha}(t)/dt$ \cite{Redner}. In our case of the
absorbing boundary condition at X we obtain the first passage density in terms of
the flux at $x=\mathrm{X}$ ($=1$ in our units). In Laplace space,
\begin{equation}
\wp_{\alpha}(s)=j(1,s)=-K_{\alpha}\alpha_2A_2e^{-\alpha_2x_{\mathrm{A}}}-
K_{\alpha}\beta_2B_2
e^{-\beta_2x_{\mathrm{A}}}
\label{flux}
\end{equation}
in terms of the exponents and coefficients defined in Eqs.~(\ref{exponents}) and
(\ref{system}).

We note that for CTRW subdiffusion any process described by the fractional
Fokker-Planck equation (\ref{ffpe}) can be related to its Brownian counterpart
simply by the method of subordination, i.e., a transformation of the number of
steps to the process time. For the first passage process this subordination
corresponds to the Laplace space rescaling \cite{report}
\begin{equation}
\wp_\alpha(s)=\wp_1\left(s^\alpha\frac{\eta_{\alpha}}{\eta_1}\right),
\label{PDF}
\end{equation}
where the factor $\eta_{\alpha}/\eta_1$ takes care of the dimensionality:
$[\eta_{\alpha}/\eta_1]=\mathrm{sec}^{\alpha}$.

From the above expressions in Laplace space we now perform a numerical Laplace
inversion \cite{LaplaceInversion} and compare the obtained results to simulations
of the CTRW process in the external, piecewise linear potential $U$.

\section{Numerical analysis and Monte Carlo simulations}

\begin{figure}
\includegraphics[width=15.9 cm]{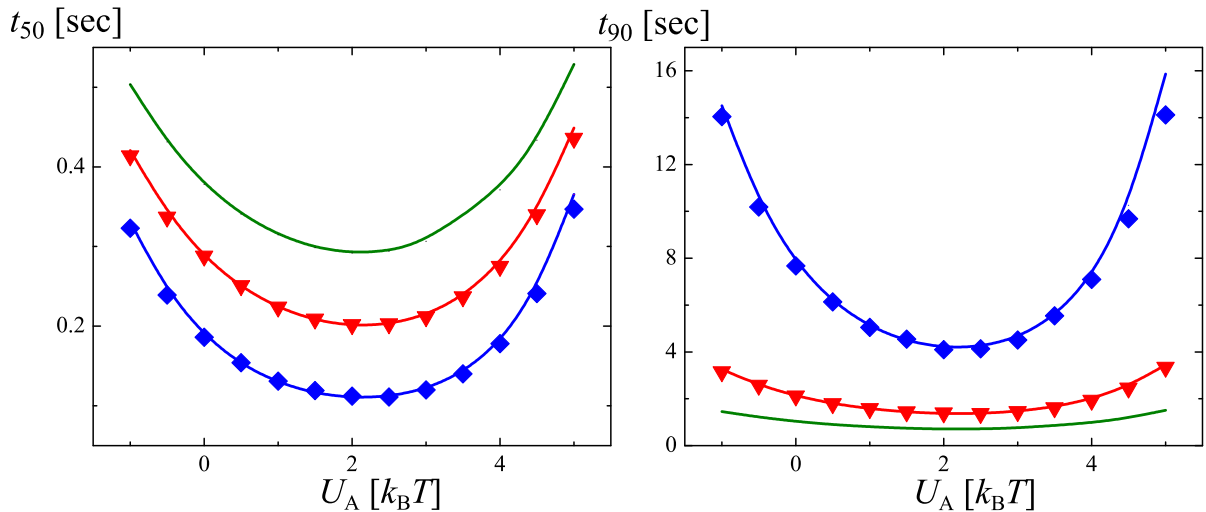}
\caption{Mean first passage time for the 50\% (left) and 90\% (right) probability
of particles having arrived to X, as function of the potential amplitudes
$U_{\mathrm{A}}$.
The green curves (top curve on the left, bottom curve on the right) correspond to
the analytical result for normal diffusion ($\alpha=1$), see Ref.~\cite{mfpt_min}.
The red (centre) curves and triangles represent the numerical Laplace inversion
of Eq.~(\ref{flux}) and simulations results for $\alpha=0.75$, respectively.
The blue curves and squares stand for $\alpha=0.5$. Lattice size for the CTRW was
$N=1001$ and $x_{\mathrm{A}}=0.1\mbox{ cm}$. Each symbol represents $10^5$
simulation runs.}
\label{t50t90}
\end{figure} 

The Monte Carlo simulations of the CTRW process were performed on a lattice with
$N=1001$ points and the waiting times in between successive jumps were drawn from
a waiting time with asymptotic power-law decay, $\psi(t)\simeq t^{-1-\alpha}$ with
$0<\alpha<1$, for details see Ref.~\cite{LevyGenerator}. In the chosen units the
length of the interval $\overline{\mathrm{OX}}$ is $1\mbox{ cm}$, and thus the
lattice constant is $\Delta x=\overline{\mathrm{OX}}/N$. For comparison with the
simulations we make use of the explicit derivation of the FFPE
\cite{report,CTRWtoFFPE}, such that in the limit $\tau^{\star}\to0$ and $\Delta x
\to0$ we have
\begin{eqnarray}
\nonumber 
K_{\alpha}\approx\frac{1}{2N^2\tau^{\alpha}},\\
\nonumber
|v_1|\approx\frac{U_{\mathrm{A}}}{2x_{\mathrm{A}}N^2k_BT_M\tau^{\alpha}},\\
|v_2|\approx\frac{U_{\mathrm{A}}}{2(1-x_{\mathrm{A}})N^2k_BT_M\tau^\alpha},
\end{eqnarray}
where $k_B$ is the Boltzmann factor and $T$ the (Monte-Carlo simulations)
temperature. For simplicity we use $K_\alpha=1\,\mathrm{cm}^2/\mathrm{sec}^\alpha$.

Results are shown in Fig.~\ref{t50t90} for the 50\% and 90\% probability of
particle absorption at the target point X. For each case we show results for
the cases $\alpha=1/2$ and $\alpha=3/4$, as well as include the analytical
result for the Brownian case from Ref.~\cite{mfpt_min}. In these simulations
the potential maximum was placed at $x_{\mathrm{A}}=0.1\mbox{ cm}$. Moreover, the potential
at the starting and end points was chosen as zero: $U_\mathrm{O}=U_{\mathrm{X}}=0$.
The lines for the subdiffusive cases were obtained from numerical Laplace inversion
of Eq.~(\ref{flux}) and subsequent integration such that the plotted times $t_{50}$
and $t_{90}$ are implicitly defined through the integral $\int_0^{t_{50}}\wp_{
\alpha}(t)dt=0.5$, and analogously $=0.9$ for $t_{90}$. The symbols are obtained
from the Monte Carlo simulations.

Fig.~\ref{t50t90} shows some remarkable properties. Thus, for the case of the 90\%
probability the Brownian case exhibits the shortest absorption times, and the
subdiffusive cases with $\alpha=3/4$ and $\alpha=1/2$ become increasingly slower.
This result would be naively expected. However, for the 50\% probability the
behaviour is exactly opposite, i.e., the 50\% first passage is fastest for the
most pronounced subdiffusion. This effect is due to the fact that one-sided
stable distributions, to which our waiting time distribution $\psi(t)$ belongs,
have long power-law tails, but are also more concentrated around the origin at
$t=0$. Thus, if we cut off extremely long waiting times governed by the long
tail of $\psi(t)$ (particles that do not arrive up to $t_{50}$), we actually
observe that the resulting process becomes faster for decreasing $\alpha$. At
90\% probability this trend is inverted, as the statistics include sufficiently
many long(er) waiting times.

\begin{figure}
\begin{center}
\includegraphics[width=10 cm]{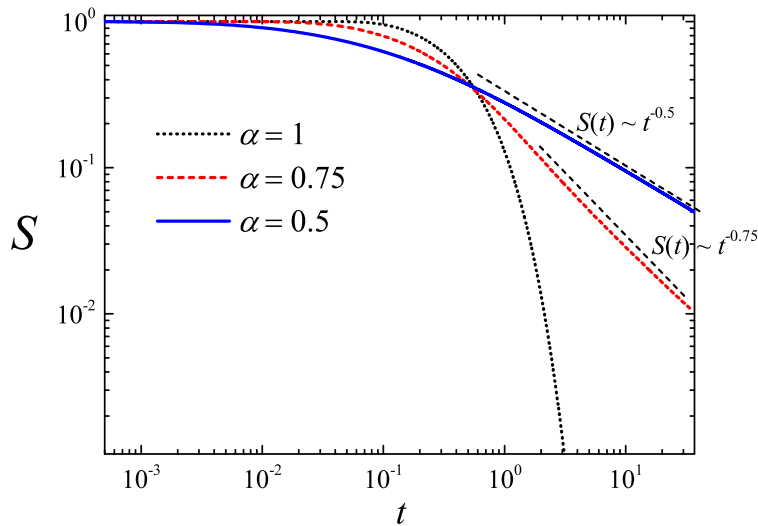}
\end{center}
\caption{Survival probability as function of time $t$ for the Brownian case and
stable exponents $\alpha=0.5$ and $\alpha=0.75$. While for $\alpha=1$ the decay
of $\mathscr{S}_{\alpha}(t)$ is exponential, for the subdiffusive cases a
power-law behaviour
is observed as $t\to\infty$. Before the crossover at $t\approx1$ the decay of the
subdiffusive particles is faster than the Brownian particle. Parameters: $U_{
\mathrm{A}}=2k_BT$ and $x_{\mathrm{A}}=0.1\mbox{ cm}$.}
\label{survival}
\end{figure} 

The second important observation is that the minima of all first passage time
curves in Fig.~\ref{t50t90} are located at $U_{\mathrm{A}}\approx2.2k_BT$ for
all $\alpha$ as well as for 50\% and 90\% probability. Thus, if a certain
value $U_{\mathrm{A}}$ optimises the first passage behaviour for a Brownian
particle, it also optimises the corresponding subdiffusive dynamics.

These findings are corroborated by the functional behaviour of the survival
probability $\mathscr{S}(t)$, as shown in Fig.~\ref{survival}. At smaller
times, corresponding to a smaller percentage of the probability of first
passage, indeed the decay is faster for more pronounced subdiffusion and
slowest for normal diffusion. Approximately at unit time $t=1$ a crossover
is observed, and for longer times we find the naively expected behaviour:
Brownian motion effects the fastest decay while the subdiffusive motion is
slower. At long times $t\rightarrow\infty$ the survival probability for the
subdiffusive cases exhibits the power law $\mathscr{S}_{\alpha}(t)\sim t^{-
\alpha}$, compare Refs.~\cite{report,physica,Lindenberg}. This follows
directly from the subordinated exponential decay of Brownian motion also shown
in Fig.~\ref{survival}.

\begin{figure}
\includegraphics[width=16 cm]{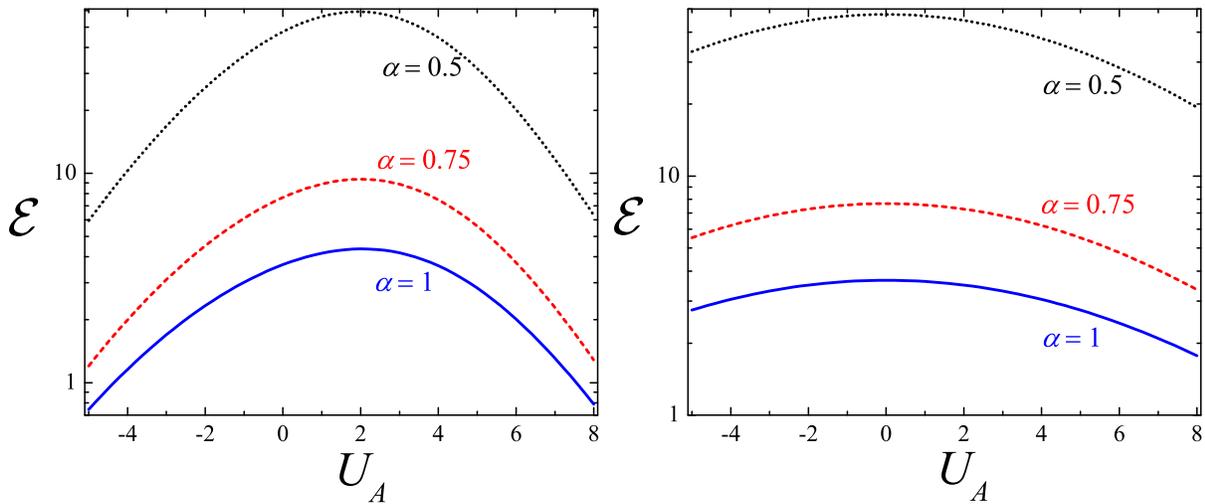}
\caption{Dependence of the efficiency $\mathcal{E}$ on the value $U_{\mathrm{A}}$
of the potential at the maximum point A, for different positions of point A,
$x_{\mathrm{A}}=0.1\mbox{ cm}$ (left, asymmetric potential) and $x_{\mathrm{A}}=0.5$ cm (right, symmetric
potential).}
\label{EffX01X05}
\end{figure} 

Yet another way to characterise the first passage behaviour is to evaluate the
average rate $\left<1/t\right>$, which was introduced for superdiffusive search
as a non-diverging measure of optimisation \cite{LevyShort,LevyLong}. This
quantity can be computed conveniently numerically via the relation
\begin{equation}
\mathcal{E}=\left\langle\frac{1}{t}\right\rangle=\int_0^{\infty}\wp_{\alpha}(s)ds
\end{equation}
from the Laplace space result of the first passage density $\wp_{\alpha}(t)$. The
functional
form of this quantity shows another remarkable property: even in the absence of a
potential the efficiency $\mathcal{E}$ increases with decreasing stable exponent
$\alpha$. Namely, using the known result for normal Brownian diffusion
\cite{Redner}, from subordination we find that for $U(x)=0$
\begin{equation}
\wp_{\alpha}(s)=2\frac{\sinh\left(L\sqrt{s^{\alpha}/D}\right)}{\sinh\left(2L\sqrt{
s^{\alpha}/D}\right)}=\frac{1}{\cosh\left(L\sqrt{s^{\alpha}/D}\right)}
\end{equation}
Therefore,
\begin{equation}
\mathcal{E}=\int_0^{\infty}\cosh\left(L\sqrt{s^{\alpha}/D}\right)ds\sim C+\int_0^{
\infty}e^{-Ls^{\alpha/2}/\sqrt D}ds,
\end{equation}
where $C$ is a constant. In the limit $\alpha\rightarrow0$ this expression reduces
to
\begin{equation}
\mathcal{E}\approx\Gamma\left(\frac{2}{\alpha}\right),
\end{equation}
which proves the fact that the average rate increases with the decrease of the
stable exponent $\alpha$. In the case of the piecewise linear potential this
behaviour is indeed preserved, as shown in Fig.~\ref{EffX01X05}. The figure
also demonstrates that the maxima of the efficiency are consistent with the
minima of the 50\% and 90\% probability first passage times, $t_{50}$ and $t_{90}$
shown in Fig.~\ref{t50t90}. Note that even in the case of a symmetric potential
shown on the right of Fig.~\ref{EffX01X05} the first passage dynamics profits
from the existence of the energy landscape: after crossing the peak the diffusion
back towards the starting point O is suppressed.

\begin{figure}
\includegraphics[width=15.8cm]{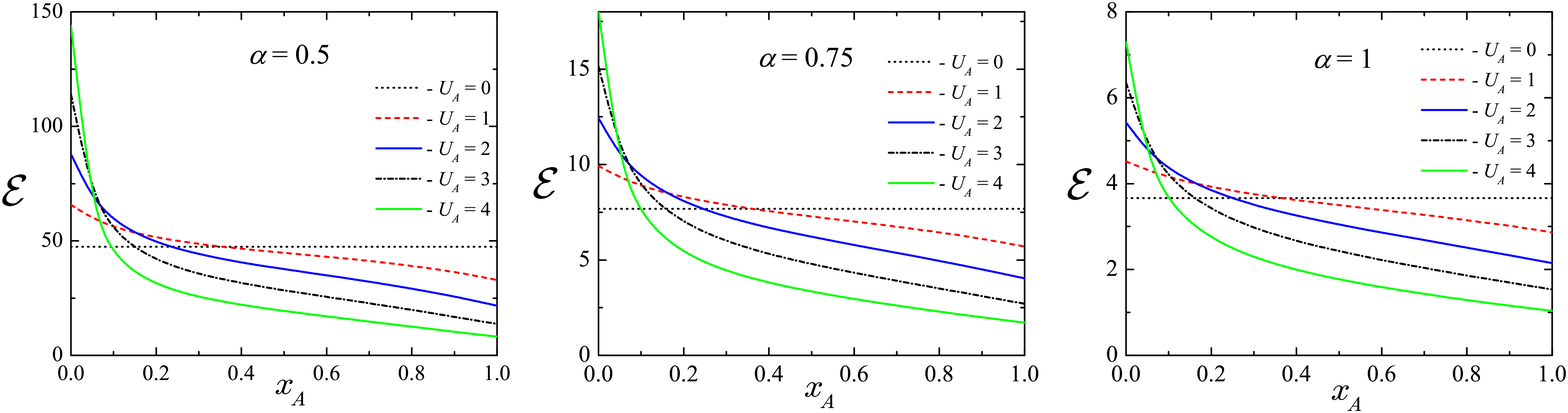}
\caption{Efficiency $\mathcal{E}$ as function of the position $x_{\mathrm{A}}$ of
the potential maximum with fixed value $U_{\mathrm{A}}$. Note the crossover at
small values of $x_{\mathrm{A}}$. A higher potential peak $U_{\mathrm{A}}$
facilitates the first passage as long as the peak location $x_{\mathrm{A}}$
remains close to the starting point O. The dotted line indicates the efficiency
value in absence of an external potential, i.e., $U_{\mathrm{A}}=0$.}
\label{EffVarXA}
\end{figure} 

The dependence of the efficiency $\mathcal{E}$ on the position $x_{\mathrm{A}}$
of the potential maximum is displayed in Fig.~\ref{EffVarXA}, for different
$U_{\mathrm{A}}$ values of the potential (as indicated in the panels). Consistently
for all $\alpha$ the efficiency increases with growing value $U_{\mathrm{A}}$ when
$x_{\mathrm{A}}$ shifts towards the starting point O: the particle requires a
larger thermal fluctuation to cross the initial peak $U_{\mathrm{A}}$, but then
experiences a higher, constant drift velocity towards its target X. At growing
values for $x_{\mathrm{A}}$ we observe a crossover, and then a higher peak value
$U_{\mathrm{A}}$ effects lower efficiency.

What would happen if we let the particles age before sending them on their
journey from O to X? Ageing is a characteristic property of subdiffusive CTRW
particles: the process is highly non-stationary, and physical observables
strongly depend on the ageing time $t_a$ elapsing between system initiation
and start of the measurement \cite{Eli1}. Due to the scale-free form of the
waiting time distribution with diverging mean waiting time $\langle t\rangle$,
longer and longer waiting times occur on average, such that effectively the
particle is constantly slowing down. Physically, in the picture of a random
energy landscape this effect corresponds to the situation that the particle
discovers deeper and deeper traps on its path. After passing the ageing period,
the first step of the particle then corresponds to the forward waiting time
\cite{Eli1,michael}
\begin{equation}
\psi_{t_a}(t_1)=\frac{\sin(\pi \alpha)}{\pi}\frac{t_a^{\alpha}}{t_1^{\alpha}
(t_1+t_a)}
\label{FirstJumpPDF}
\end{equation}

\begin{figure}
\begin{center}
\includegraphics[width=16 cm]{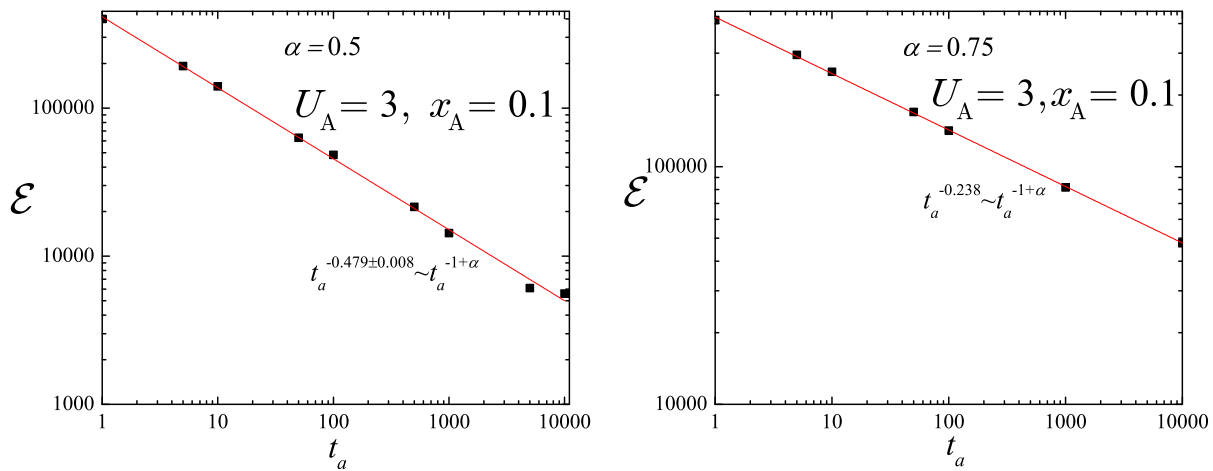}
\end{center}
\caption{Dependence of the search efficiency $\mathcal{E}$ on the ageing time $t
_a$ elapsing between initiation of the process and start of our observation of
the particle at $x_{\mathrm{O}}$, for $\alpha=0.5$ (left) and $\alpha=0.75$
(right). We observe that $\mathcal{E}(t_a)\simeq t_a^{\alpha-1}$, as predicted
by Eq.~(\ref{E_age}). Parameters: $U_{\mathrm{A}}=3$ and $x_{\mathrm{A}}=0.1$.}
\label{Aging}
\end{figure}

For the parameters $U_{\mathrm{A}}=3$ and $x_{\mathrm{A}}=0.1$ we show the dependence of the efficiency
$\mathcal{E}$ on the ageing time $t_a$ of the process for the subdiffusive cases
with $\alpha=1/2$ and $\alpha=3/4$. Indeed, the continued slow-down of the motion
due to increase of the typical waiting times leads to a pronounced decrease of
$\mathcal{E}$ of the power-law form
\begin{equation}
\label{E_age}
\mathcal{E}(t_a)\simeq t_a^{\alpha-1},
\end{equation}
corresponding to the product of the first passage density without ageing ($t_a=0$)
and the probability density of the forward waiting time \cite{Eli1}. This behaviour
is indeed nicely observed in our simulations, as demonstrated in Fig.~\ref{Aging}.

\section{Discussion and Conclusions}

The study of the mean first passage behaviour in finite, flat potential landscapes
has experienced remarkable progress during the last few years \cite{olivier1}. In
particular, the distinction of compact versus non-compact exploration strategies
led to the concept of ``geometry-controlled kinetics'' \cite{olivier2}. Moreover,
the trajectory-to-trajectory variation of first passage times have been studied
on finite domains of various shapes recently \cite{gleb}. Much less is known
about the first passage in potential landscapes.

Previously we found that for normal Brownian diffusion the first passage in a
finite interval can be sped up significantly by introducing a potential landscape
\cite{mfpt_min}. For the case of a piecewise linear potential, the role of the
potential is intuitively clear: after crossing an initial potential barrier, the
particle continuously slides down towards the target. When the position of the
barrier successively approaches the point of release and the barrier height
increases, the mean first passage time decreases \cite{mfpt_min}.

Here we studied the behaviour in such a piecewise linear potential for the case
of a subdiffusing particle whose dynamics is governed by a long-tailed waiting
time distribution with a diverging characteristic waiting time. From analysis of
the probability percentage of the first passage, the survival probability, and
the efficiency parameter we found a number of remarkable properties. Namely,
as in the Brownian case the introduction of the barrier indeed leads to a more
efficient (i.e., faster) first passage behaviour for a given value $\alpha$ of
the anomalous diffusion exponent. At short times, somewhat surprisingly the more
subdiffusive particle performs better than the Brownian particle,
while at longer times the less subdiffusive particle wins out.
The optimal height of the potential barrier is thereby conserved for varying
$\alpha$. Moreover, the efficiency increases dramatically when the position of
the potential maximum is shifted towards the particle origin O. Finally, the
ageing of the particle leads to a power-law decay of the efficiency as function
of the ageing time.

As in the Brownian case \cite{mfpt_min} a faster first passage of particles through
the interval $\overline{\mathrm{OX}}$ in the presence of the potential barrier
does not contradict the corresponding result of the (fractional) Kramers escape
\cite{AnomalousKramers}. The distribution of escape times for CTRW subdiffusion
becomes \cite{AnomalousKramers}
\begin{equation}
\wp^K_{\alpha}(t)=E_\alpha\left(-r_K^{(\alpha)}t^{\alpha}\right)
\label{rate}
\end{equation}
where $E_\alpha$ is the Mittag-Leffler function
\begin{equation}
E_{\alpha}(-z)=\sum_{n=0}^{\infty}(-z)^n/\Gamma(1+\alpha n)
\end{equation}
with $E_{\alpha}(-z)\sim1/z$ at $z\to\infty$. The (fractional) rate coefficient
$r_K^{(\alpha)}$ is given by
\begin{equation}
r_K^{(\alpha)}=\frac{\sqrt{U''(x_{\mathrm{min}})U''(x_{\mathrm{max}})}}{2\pi
m\eta_\alpha}\exp\left(-\frac{\Delta U}{k_B T}\right).
\label{Kramers}
\end{equation} 
Thus shifting the maximum at $x_{\mathrm{A}}$ towards point O leads to an increase
of the curvatures at the maximum and minimum positions in Eq.~(\ref{Kramers}), and
hence speeds up the rate in the same way as observed in Ref.~\cite{mfpt_min}, if
$U_{\mathrm{A}}$ is kept constant.

Let us briefly discuss a technologically relevant, physical application of the
above results. Namely, we consider the passage of a polymer chain through a
narrow channel, the so-called translocation process \cite{meller}. Indeed, in
terms of the translocation co-ordinate $m$ (the number of monomers crossing the
exit of the channel) the translocation becomes subdiffusive, see, for instance,
Refs.~\cite{KantorKardar,epl}. The fractional Fokker-Planck equation was proposed
to model this subdiffusive behaviour, and shown to capture the first passage
behaviour of the translocating polymer chain in comparison to simulations
\cite{Dubbeldam}. More recently, it has become clear that the translocation of
a polymer through a narrow channel without interactions between channel wall
and polymer chain is stochastically described by fractional Langevin equation
motion driven by long-ranged Gaussian noise \cite{fle}.
However, the long-time translocation
dynamics of a polymer may still be governed by the fractional Fokker-Planck
equation if the motion of the chain is successively immobilised with power-law
distributed waiting times due to monomer-channel interactions or extra-channel
inhibitors.

The translocation time of a polymer through a channel consists of three distinct
contributions \cite{Muthu2010}: (i) the time needed for the free polymer chain on
the cis side of the channel to diffuse to the channel entrance, (ii) the time for
one of the chain ends to thread into the channel entrance, and (iii) the
passage of the chain through the channel across the entropic potential describing
the reduction of the polymer's accessible degrees of freedom due to the imposed
constraints plus some external driving potential. With the translocation co-ordinate
$m$ for a polymer with $N$ monomers, this gives rise to the free energy landscape
\cite{Muthu1999}
\begin{equation}
\mathcal{F}(m)=-k_BT\left(N\ln\mu+\left(\gamma_1-1\right)\ln\left[(N-m)m\right]
\right),
\label{entropy}
\end{equation}
where $\mu$ is the (non-universal) lattice connectivity (e.g., $\mu=6$ on a cubic
lattice) and $\gamma_1$ is the topological critical exponent for a self-avoiding
chain attached to a wall, $\gamma_1=0.680$ for a self-avoiding chain in three
dimensions. For translocation processes only the second term of the free energy
function (\ref{entropy}) depends on the translocation co-ordinate $m$ and thus
matters to our analysis. The entropic free energy barrier in Eq.~(\ref{entropy})
clearly slows down the translocation of the chain. In the mathematical description
in the pseudo-equilibrium approximation \cite{Muthu2010,Muthu1999,meluo}, we
are only interested in the very translocation dynamics of the chain, and we
therefore impose a reflecting boundary condition at $x=0$, in order to prevent
the chain from escaping the channel.

\begin{figure}
\begin{center}
\includegraphics[width=8.6 cm]{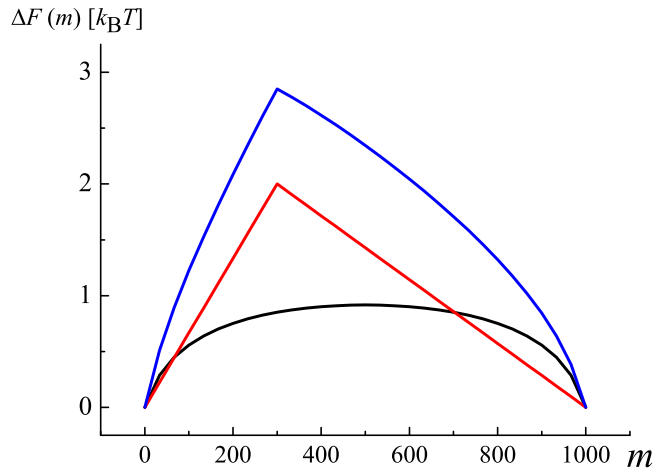}
\end{center}
\caption{Potential landscape for translocation of a gradient copolymer in the
quasi-equilibrium approximation. The black line depicts the entropic contribution
of the polymer chain. The red line is the piecewise linear potential, which
characterises the interactions, see text. The blue curve is
the sum of these contributions which represents the actual potential for the
translocation process.}
\label{LandscapeTrans}
\end{figure} 

What happens when instead of a homogeneous polymer chain we consider a gradient
or tapered copolymer with a a sequence of monomers of different types with
different monomer-channel interactions? Gradient copolymers show a quite special
behaviour with respect to their thermodymamic properties, contrasting other
copolymer sequences \cite{Beginn,Gradient2012,JCP2007,GradMicelle}.
The effects of interactions between translocating chain and channel were
indeed studied previously for the passage of heteropolymers
Refs.~\cite{golestanian}. To
demonstrate the possibility of modifying the translocation time statistics for
such inhomogeneous polymer chains we imagine a polymer sequence, that gives rise
to a piecewise linear potential of the shape studied above. Combining this with
the translocation free energy (\ref{entropy}), we obtain the landscape portrayed
in Fig.~\ref{LandscapeTrans}.

\begin{figure}
\begin{center}
\includegraphics[width=16 cm]{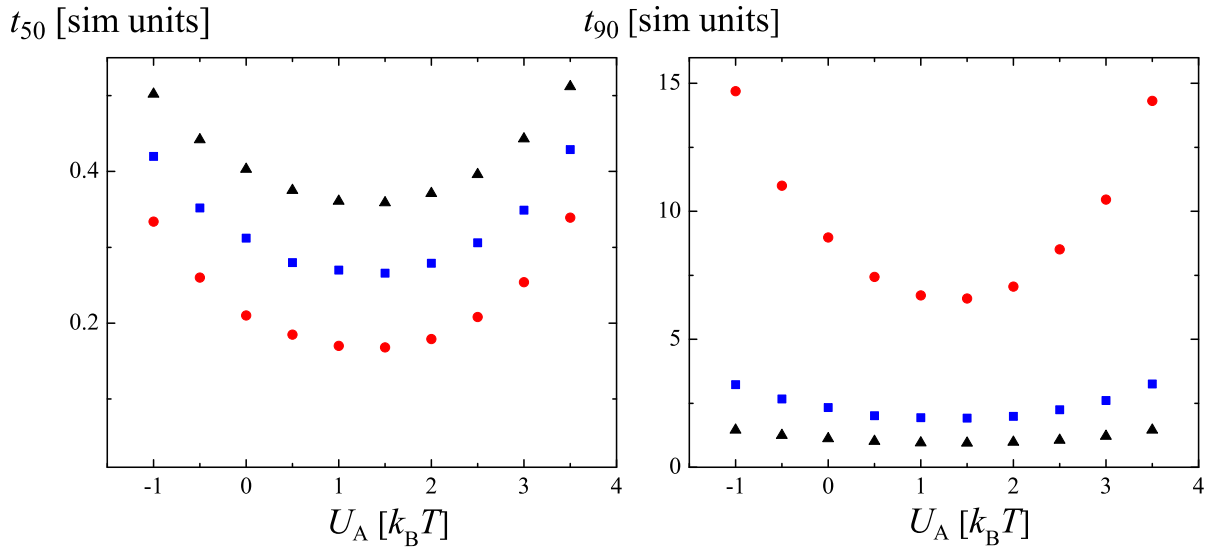}
\end{center}
\caption{Simulated translocation times with 50\% and 90\% translocation probability
for the combined free energy landscape of Fig.~\ref{LandscapeTrans} for a chain
with $N=1001$ monomers. The maximum of the piecewise linear potential component
is at $x_{\mathrm{A}}=100$. Black triangles show the simulations results for
$\alpha=0.5$, blue squares for $\alpha=0.75$, and red circles for the Brownian
case. $10^5$ simulation runs were performed to obtain the data points.}
\label{translocation}
\end{figure}

Simulations of this modified translocation process produce the 50\% and 90\%
probabilities for translocation across the combined free energy landscape shown
in Fig.~\ref{translocation}. Evidently the piecewise linear potential can lower the
translocation time. As above, the position of the optimum (at around $U_\mathrm{A}
\approx1.5k_BT$, different from the value without the entropic potential) is
independent of the exponent $\alpha$ and whether we consider the 50\% or 90\%
case. Notably, while the overall translocation times are higher in the 90\% case,
the effect of the potential is strongest for $\alpha=1/2$.

Thus the sequence of the copolymer may have a significant influence on the
translocation times. This statement itself was already confirmed by Monte-Carlo
simulations Ref.~\cite{Slater2008}, in which the authors found a distinct
variation of up to 3 orders of magnitude in translocation times for different
sequences but same composition. However, they did not report the effect of
decreased translocation times in comparison with the homopolymer by adding
slower translocating segments. We mention that an essential part of this
brief discussion relies on the quasi-equilibrium approximation, necessary to
apply a free energy picture \cite{KantorKardar}, which was shown to be
inapplicable in many cases of translocation \cite{KantorKardar,Sakaue}. However,
we assume that these effects on polymer translocation times will remain relevant
for non-equilibrium situations, as well. It is feasible that parts of biopolymer
sequences may have gradient or tapered parts which assist translocation. These
specific interaction potentials may also originate from combination of complex
pore structure and copolymer sequence as in Refs.~\cite{golestanian}. Further
studies of this problem are expected to shed new light on the biophysics of
translocation processes.

\ack

VVP acknowledges financial support from the Deutsche Forschungsgemeinschaft
(project PA2042/1-1) as well as discussions with J. Schulz about the simulation
of random variables. RM acknowledges support from the Academy of Finland within
the FiDiPro scheme.

\end{document}